\documentstyle[prl,twocolumn,aps]{revtex}
\begin{document}
\draft
\title{Regime interpretation of anomalous vortex dynamics in 2D superconductors}
\author{Massimiliano Capezzali$^{(a)}$, Hans Beck$^{(a)}$ and Subodh R.
Shenoy$^{(b)}$}
\address{$^{(a)}$ Institut de Physique, Universit\'{e} de Neuch\^{a}tel, Rue A.L.
Breguet 1,
2000 Neuch\^{a}tel, Switzerland \\
$^{(b)}$ Condensed Matter Physics Group, International Centre for Theoretical
Physics, 34100 Trieste, Italy}
%
\twocolumn[
\maketitle
\widetext

\vspace*{-1.0truecm}

\begin{abstract}
\begin{center}
\parbox{14cm}{Low -frequency dynamic impedance
($\sigma^{-1}(\omega,T)\equiv(\sigma_{1}+i\sigma_{2})^{-1}$)
measurements on Josephson junction arrays with finite vortex screening length
$\xi$, found that
$\sigma_{1}\sim\left|\log{\omega}\right|$, $\sigma_{2}\sim$ constant.
This implies anomalously sluggish vortex mobilities 
$\mu_{V}(\omega)\sim\sigma_{1}^{-1}$, and is in conflict with general dynamical scaling
expressions that yield, for low-$\omega$, $\sigma_{1}\rightarrow\xi^{2}$ and 
$\sigma_{2}\rightarrow 0$. 
We calculate : a) $\sigma(\omega,T)$
by real-space vortex scaling;
b) $\mu_{V}(\omega)$ using Mori's formalism for a screened Coulomb gas.
We find, in addition to the usual critical (large-$\omega$) and hydrodynamic
(low-$\omega$) regimes, a new intermediate-frequency scaling regime into which the
experimental data fall. This resolves the above mentioned conflict and makes explicit 
predictions for the scaling form of $\sigma(\omega,T)$, testable in SNS and SIS arrays.}
\end{center}
\end{abstract}
\pacs{
\hspace{1.9cm}
}]
\narrowtext
The dynamic conductivity of superconductors 
$\sigma(\omega,T)\equiv\sigma_{1}+i\sigma_{2}\equiv
|\sigma|e^{i\phi_{\sigma}}$, including high-$T_{C}$ materials and Josephson junction
arrays (JJA), has been the focus of much recent interest 
\cite{fisher,dorsey,minnhagen2,theron}. Dynamical scaling forms, $\sigma=
\xi^{2+z-d}S_{\pm}(Y^{-1})$, $\phi_{\sigma}=\Phi(Y^{-1})$ were proposed by Fisher et al. 
\cite{fisher} and Dorsey \cite{dorsey}, where $z$ is the dynamic exponent, and 
$Y\sim 1/\omega\xi^{z}$. The results apply \cite{fisher,dorsey} also for the vortex-
unbinding Kosterlitz-Thouless (KT) transition in $d=2$ \cite{thouless}, 
$\xi$ being the vortex screening
length. The scaling functions $\Phi_{\sigma}$ and $S_{+}$ ($S_{-}$) have
well-defined limits in the hydrodynamic ($Y\gg 1$) and critical ($Y\ll 1$) regimes, e.g. 
$\sigma_{1}\rightarrow\xi^{2}$, $\sigma_{2}\rightarrow 0$ in the
$Y\rightarrow\infty$, dc limit. 2D JJA's \cite{mooij} are clean, controllable 2D superconductors, and should be
ideal systems to display the universal dynamic scaling behaviour and limits.\\ 
Remarkably, however, 
low-frequency dynamic impedance $(\sigma^{-1})$ measurements
\cite{theron} on 2D SNS triangular-lattice JJA's with a (field-induced) vortex screening
length $\xi$, find $\sigma_{1}\sim|\ln{\omega}|$, $\sigma_{2}\sim$ constant, in conflict
with dynamic scaling limits. This also implies anomalously sluggish 
low-frequency vortex mobilities,
$\mu_{V}(\omega)\sim\sigma_{1}^{-1}\sim 1/|\ln{\omega}|$
going to zero for $\omega\rightarrow 0$. 
$\sigma$ is related to the dynamic dielectric function $\epsilon(\omega)$ :
$\sigma\sim iY/\epsilon(\omega)$. Surprisingly, Minnhagen's phenomenology (MP)
for $\epsilon(\omega)$, described below, and related simulations
support
this anomalous behaviour, but understanding the apparent
breakdown of dynamical scaling, in the very arena where one might
expect its clear verification, is of central importance.\\
In this Letter, we reconcile these results, by a
"regime interpretation" \cite{capezzali} defined by the ratio $Y=(r_{\omega}/\xi)^{2}$
of the (squares of the) frequency-dependent diffusive probe length \cite{ahns}
 $r_{\omega}=\sqrt{\Gamma_{0}/\omega} $ and the screening length
$\xi$. Here, $\Gamma_{0}$ is a junction-determined phase diffusion
rate, the lattice constant is unity, and we consider weak screening
and probes over several lattice constants : $\xi\gg1$, $r_{\omega}\gg1$.
We a) recalculate $\sigma(\omega,T)$ by a real-space
scaling \cite{shenoy}, with an improved treatment of
intermediate-scale screening; b) evaluate the vortex mobility
$\mu_{V}$ using Mori's formalism for a screened Coulomb gas \cite{beck}.
Three probe-scale regimes emerge.
I) Probing free-vortex scales ("low" frequencies) $r_{\omega}\gg\xi$,
in a "hydrodynamic" region, Drude behaviour with the correct \cite{fisher,dorsey} dc 
conductivity,
$\sigma_{1}(\omega\rightarrow0,T)\sim\xi^{2}$, is recovered. II)
At intermediate scales/frequencies, 
$r_{\omega}{\ \lower-1.2pt\vbox{\hbox{\rlap{$<$}\lower5pt\vbox{\hbox{$\sim$}}}}\ }\xi$, 
in a new "precritical"
region,
MP-like behaviour, $\sigma_{1}\sim\left|\log{\omega}\right|$ is found.
III) Probing bound pair-scales ("high" frequencies)  $r_{\omega}\ll\xi$
in a "critical" region extending from just above $T_{KT}$ to $T=0$, a
scale-dependent vortex damping
$\sim\sigma_{1}\sim(r_{\omega}/\ln{\omega})^{2}
\sim(\omega(\ln{\omega})^{2})^{-1}$
is found, corresponding to large pairs moving in a logarithmically interacting
viscous medium of smaller pairs. The results of $\omega\rightarrow0$,
$T \rightarrow T_{KT}^{+}$, thus depend on the order of the limits. 
The ratio $R_{\sigma}=\sigma_{1}(\omega,T)/\sigma_{2}(\omega,T)
\equiv\cot{\phi_{\sigma}}$ at $Y=1$
interpolates between Drude ($R_{\sigma}=1$) and MP ($R_{\sigma}=2/\pi$)
signatures \cite{minnhagen}, as $\omega$ increases from zero, 
or $T$ increases from  $T_{KT}^{+}$.
As a satisfying byproduct of calculation a), the MP-like expressions emerge
as approximations 
to the $\sigma/\xi^{2}$ scaling function, valid in regime II. Calculation b) demonstrates that the
general results are independent of the details of JJA dynamics, and
depend only on Coulomb gas screening properties. 
Local spin-wave damping mechanisms specific to SNS arrays \cite{beck,korshunov}
could play an additional role in producing anomalous behaviour, widening regime II.
But both SNS and
SIS arrays should show all three regimes in principle, with different
relative sizes of regimes I and II, coming from very general considerations. \\
\indent
Two different physical circumstances yield a non-zero free vortex density, and
thus finite $\xi$ : i) For zero external flux, and $T>T_{KT}$, $\xi^{-1}=\xi_{+}^{-1}(T)
\sim e^{-\left( T-T_{KT}\right)^{-1/2}}\neq 0$ above transition, $T>T_{KT}$; while 
$\xi^{-1}=0$ for $T<T_{KT}$. ii) Flux-induced vortices of concentration $f\ll 1$
\cite{theron} (too dilute to form a stable lattice) can form a one-component plasma with a
screening length $\xi$, given by the
Debye expression $\xi^{-1}=\xi^{-1}_{D}(f)=(4\pi^{2}f/\bar{T})^{1/2}\neq 0$,
corresponding to "above transition" for any $T$.\\
We now sketch the MP ideas \cite{minnhagen}, originally developed to
describe $\sigma(\omega,T)$ structures at $T=T_{\omega}>T_{KT}$, where
$\xi_{+}(T_{\omega})=r_{\omega}$. The
zero-wavevector conductivity $\sigma(\omega,T)$
 is proportional to the corresponding (inverse) dielectric constant:
$\sigma(\omega,T)/\sigma_{0}K_{0}\xi^{2}=
i Y\left[\varepsilon_{V}(k=0,\omega,T)\right]^{-1}$, 
where $\sigma_{0}$ is a conductivity scale and $K_{0}$ the bare
vortex coupling. In MP, the real part 
$\Re\left(\varepsilon_{V}(k=0,\omega,T)^{-1} \right)$
of this zero-wave-vector dynamic function is
approximated by the zero-frequency static function,
$\Re\left(\varepsilon_{V}(k,\omega=0,T)^{-1} \right)$,
evaluated at the probe scale, $k=r_{\omega}^{-1}$.
The imaginary part, $\Im\left(\varepsilon_{V}(k=0,\omega,T)^{-1} \right)$
is found from the Kramers-Kronig (KK) relations, that produce a $\ln{Y}$ dependence. 
Thus \cite{minnhagen}, 
with $\varepsilon_{V}(k,\omega=0,T)^{-1}=\vec{k}^{2}/(\vec{k}^{2}+\xi^{-2})$ :
\begin{equation}
{\sigma_{2} \over \sigma_{0}K_{0}\xi^{2}}={Y \over Y+1} ,~~~~~~{\sigma_{1}
\over \sigma_{0}K_{0}\xi^{2}}={2 \over \pi}{Y^{2}\ln{Y} \over Y^{2}-1} .
\end{equation}
At $Y=1$, $R_{\sigma}=2/\pi$ (i.e. $\phi_{\sigma}=\arctan{(\pi/2)}$), 
an MP signature. 
The dynamical scaling limits \cite{fisher,dorsey} in the $Y\ll 1$ critical regime, both above
and below $T_{KT}$, are $\sigma_{1}\sim\sigma_{2}\sim1/\omega$ (independent of $\xi$),
$ \Phi_{\sigma}=\pi/2$. 
In the $Y\gg 1$ hydrodynamic regime, for $T>T_{KT}$
($T<T_{KT}$), 
one finds $\sigma_{1}\sim\xi^{2}$, $\sigma_{2}\sim 0$, $\Phi_{\sigma}\sim 0$
($\sigma_{1}\sim\delta(\omega)$, $\sigma_{2}\sim 1/\omega$). Eqn. (1) for MP has very
different limits, however : $\sigma_{1}\sim\xi^{2}Y^{2}|\ln{Y}|$, $\sigma_{2}\sim
1/\omega$ for $Y\ll 1$; and $\sigma_{1}\sim\xi^{2}|\ln{Y}|$, $\sigma_{2}\sim\xi^{2}$ for
$Y\gg 1$. (Note that the $Y\gg 1$ limit, for $\xi=\xi_{+}(T)$ fixed, $\omega\rightarrow 0$, implies
infinite dc conductivity, \underline{above} $T_{KT}$).
We now
outline our two complementary calculations, with details elsewhere
\cite{capezzali}, showing that MP-like behaviour emerges in an intermediate regime II,
rather than in the scaling form regimes I, III. 
\section*{(a) Real-space vortex scaling and $\sigma(\omega,T)$}
The total (dimensionless) JJA bond current $I_{\mu i}^{tot}$ ($\mu$=$x$, $y$ directions)
is a sum of Josephson or super- ($\sim\sin{\triangle_{\mu}\theta_{i}}$),
phase-slip or normal- ($\sim \triangle_{\mu}\dot{\theta}_{i}$) and
noise- currents, and is conserved at every 2D lattice site $i$. If
we ignore capacitive charge build-up on grains,
\begin{eqnarray}
\sum_{\mu}^{}{\triangle_{\mu} I_{\mu i}^{tot}} =
\sum_{\mu}^{}{}\triangle_{\mu}[
\bar{T}^{-1}\sin{(\triangle_{\mu}\theta_{i}-\dot{A}_{\mu i}(t))}+ \nonumber \\
\nu_{0}^{-1}(\triangle_{\mu}\dot{\theta_{i}}-\dot{A}_{\mu i}(t))+f_{\mu i}(t)]= 0.
\end{eqnarray}
Here,
$\nu_{0}\equiv(2eR_{J}I_{J}/\hbar)\bar{T}\equiv \Gamma_{0}\bar{T}$,
$\bar{T}^{-1}\equiv(\hbar I_{J}/2ek_{B}T)$, and $I_{J}$, $R_{J}$ are
 the junction critical current and RSJ model effective shunt resistance
 \cite{shenoy,note3}, for the SNS/SIS array. The random noise current
obeys $\left\langle f_{\mu r}(t)f_{\mu'r'}(t')
\right\rangle=(2/\nu_{0})\delta_{\mu\mu'}\delta_{rr'}\delta(t-t')$. 
The JJA grain phases are $-\pi<\theta_{i}\leq\pi$, and the external transverse vector
potential
$A_{\mu i}(t)=A_{\mu i}(\omega)e^{-i \omega t}$ is weak. 
Inverting the Laplacian $\vec{\triangle}^{2}$ on
$\dot{\theta}_{i}$, 
the Langevin dynamics equation for the phase is \cite{shenoy,teitel} :
\begin{equation}
\dot{\theta}_{r}=-\sum_{r'}^{ }{\tilde{G}_{rr'}\left[ \nu_{0}{\partial \beta
H\over \partial \theta_{r}}+\hat{F}_{r'}(t)\right]},
\end{equation}
where
$\beta H=-{1 \over \tilde{T}}\sum_{\mu,r}^{
}{\cos{(\triangle_{\mu}\theta_{r}-A_{\mu}(t))}}$,
$\tilde{G} _{rr'}=G_{rr'}-G_{rr}$ is the 2D lattice Green's function (with
singular part subtracted),
and $\left\langle \hat{F}_{r}(t)\hat{F}
_{r'}(t')\right\rangle=2\nu_{0}\tilde{G}_{rr'}\delta(t-t')$.\\
The dynamic conductivity calculation \cite{shenoy} yields
$\sigma_{\mu r,\mu'r'}\equiv\bar{\sigma}_{\mu r,\mu'r'}+\tilde{\sigma}_{\mu
r,\mu'r'}$.
Here, $\bar{\sigma}$ is the usual superfluid response, that at long wavelengths
is
$(\bar{\sigma}/\sigma_{0})=
\pi K_{\infty}\Gamma_{0}\delta(\omega)+iK_{\infty}(\Gamma_{0}/\omega)$. With
$\left\langle  \right\rangle_{0}$ denoting an average with weight
$P_{0}=e^{-\beta H (\triangle\theta-A(t))}$, $\tilde{\sigma}$ can be written as
:
\begin{equation}
{\tilde{\sigma}_{\mu r,\mu'r'} \over \sigma_{0}}\sim
\bar{T}^{-2}\int_{0}^{\infty}{e^{i \omega t}\left\langle
\sin{\triangle_{\mu}\theta_{r}}
e^{-\hat{L}_{0}t}\sin{\triangle_{\mu'}\theta_{r'}} \right\rangle_{0}} dt.
\end{equation}
As before \cite{shenoy}, we extract vortices by a dual transform,
do a gaussian truncation on spin waves, and find that the effect of the Fokker-Planck
"propagator" $e^{-\hat{L}_{0} t}$ is to produce a correlation angle decay
as  $e^{-\Gamma_{0} t}$. The correlation $\tilde{\sigma}_{\mu r,\mu'r'}$ can
be expressed as the projection (through derivatives) of the vortex partition
"generating function", with separated "test charges" at $\mu r$ and $\mu' r'$. Doing
the time integral in Eqn. (4), the $e^{-\Gamma_{0} t}$ factors lead to a
"test-charge" $(i \omega /\Gamma_{0})/(1-i \omega /\Gamma_{0})$ at
$\mu' r'$ in the partition function. 
The logarithmic potential, $\vec{\Delta}^{2}U_{0}(R/a)=+2\pi \delta_{\vec{R}, \vec{0}}$ and
dipolar ($\xi^{-1}=0$) scaling equations \cite{thouless} can be generalized 
\cite{capezzali} to include weak ($\xi^{-1}\ll 1$) monopolar screening of
$a$, $a+da$ dipole-binding, approximated by a potential
$\vec{\Delta}^{2} U(R/a)=+2\pi g_{l}\delta_{\vec{R}, \vec{0}}$. 
Real-space integration of pairs of
separation $a$, $a+da$ can then be done, as usual \cite{thouless},
producing the renormalized coupling $K_{l}$ that obeys KT scaling
equations. Since vortex damping is across the junctions in the array,
it is scale-dependent, $\Gamma_{0}\rightarrow\Gamma_{l}\equiv\Gamma_{0}/a^{2}$
($z=2$),
where
 $a \equiv e^{l}$, so the frequency-dependent test charge is
 $(i \omega a^{2}/\Gamma_{0})/(1-i \omega a^{2} /\Gamma_{0})$ \cite{note3}.
After projection, this provides a dynamic Drude factor, at scale $a\equiv e^{l}$, that
weights the incremental, (static) scaling contributions, $d(K_{l}g_{l})$. The KK relations
are thus automatically satisfied. \\
With a partial integration, the long-wavelength conductivity $\tilde{\sigma}(\omega)$ is
then, finally, an integral over all pair contributions, with a range of length
(and time) scales :
\begin{equation}
{\tilde{\sigma}(\omega) \over \sigma_{0}\xi^{2}}=
Y\int_{0}^{\infty}{dlK_{l}g_{l}\left[ {d \over dl}{{(a/r_{\omega})^{2} \over
1-i(a/r_{\omega})^{2}}} \right]} .
\end{equation}
For $a\ll 1$, $g_{l}\simeq 1$, and for
$a\gg 1$, $g_{l}$ is the Debye dielectric constant $q^{2}/(q^{2}+\xi^{2})$ at a scale
$q\sim a^{-1}$ : $g_{l}\simeq (1+(a/\xi)^{2})^{-1}$. The dominant monopole effect on
$\sigma(\omega)$ is the explicit $g_{l}$ factor, representing
a scale-dependent reduction of far-off fields, as
seen by $a$, $a+da$ dipoles. There is a smooth cross-over from dipolar
($a\ll\xi$, regime III) to monopolar ($a\gg\xi$, regime I) screening with probes 
$r_{\omega}\sim a$, and with mutual ($a\sim\xi$, regime II) screening in between. Previously,
we had matched regime I/III behaviour directly \cite{shenoy}, effectively taking
$g_{l}$ to be a step function, and suppressing the
intermediate regime.
For 
external free-vortex screening ($\xi=\xi_{D}(f)$), 
$g_{l}=1/(1+(a/\xi)^{2})$ throughout.\\
Changing variables in Eqn. (5), $a/\xi \rightarrow a$ we see
$\tilde{\sigma}(\omega,T)$ is a function of $Y$, with only logarithmic
deviations $\sim l_{\xi}\equiv\ln{\xi}$, from the limits of the integral and
 in $K_{l} \rightarrow K_{l+l_{\xi}}$. For  $\xi\propto\xi_{D}(f) \sim \sqrt{f} $,
 this implies a quasi-universality \cite {fisher} in $Y\sim f/\omega$, as found \cite{theron}.
The imaginary part $\tilde{\sigma}_{2}$ of Eqn. (5) has a function peaked at
$a^{2}/\xi^{2}=Y$ in square brackets, multiplying a roll-off function.
By rapid roll-off and sharp-peaking estimates, the \underline{total}
$\sigma_{2}$
is estimated as ($l_{\omega}\equiv \ln{r_{\omega}}$) :
\begin{equation}
{\sigma_{2}\over \sigma_{0}\xi^{2}}\approx Y\left[K_{l\xi}{Y^{-2}\over
1+Y^{-2}} , K_{l \omega}{ Y^{-1}\over 1+Y^{-1}},K_{l \omega}\right] ,
\end{equation}
in the regimes I, II, III respectively, or $Y\gg1$, $Y\leq1$, $Y\ll1$.
Note in regime III, the $\xi^{2}$ factor cancels, and \cite{shenoy}
$\sigma_{2}/\sigma_{0}\approx K_{l \omega}/\omega$ with
the correct superfluid kinetic inductance limit $K_{\infty}/\omega$,
 for $T<T_{KT}$, $\omega\rightarrow 0$. The real part of the total conductivity, 
apart from the $\delta(\omega)$ term,
(using KK relations in regime II, where the integral is harder to estimate) is :
\begin{equation}
{\sigma_{1}\over \sigma_{0}\xi^{2}}
\approx\left[ {K_{l \xi} \over 1+Y^{-2}}, {2 \over \pi}{K_{l \omega}\ln{Y} \over 1-Y^{-2}},
-{1 \over 2}{dK_{l \omega} \over dl_{\omega}}{\arctan{Y^{-1}} \over
Y^{-1}}\right]
\end{equation}
in regimes I, II, III respectively. 
The $\sigma/\xi^{2}$ results of Eqns. (6) and (7) agree with the scaling limits  
\cite{fisher,dorsey} of $S_{\pm}(Y^{-1})$ and $\phi_{\sigma}(Y)$ 
in regimes I, III. In regime II, 
with dipolar screening neglected ($K_{l\omega}\rightarrow
K_{0}$), $\sigma/\xi^{2}$
is of the MP form, Eqn. (1), . With \cite{thouless} $K_{l}\sim K_{\infty}+l^{-1}$ in regime III, 
$\sigma_{1}/\sigma_{0}\sim (\Gamma_{0}/\omega)/l_{\omega}^{2}$ for all 
$T<T_{KT}$ reflecting the KT "critical line", in the dynamics;
the phase angle \cite{fisher,dorsey} $\phi_{\sigma}=\arctan{(\sigma_{2}/\sigma_{1})}
\rightarrow\pi/2$, as $\omega\rightarrow 0$.
\section*{(b) Coulomb gas vortex dynamics}
It is important to directly calculate the Coulomb gas vortex "charge" mobility
 \cite{beck} (for $\xi^{-1}\neq 0$) in a way that is manifestly independent of
 the details \cite{note3} of JJA dynamics, but shows the two frequency regimes
I and II (since dipolar screening is not included, regime III will not appear). The
overdamped equation of motion for a charge +1 is
$\Gamma_{V}\dot{\vec{R_{l}}}=
-\sum_{j\neq i}^{}{e_{j}\vec{\nabla}V(\vec{R}_{i}-\vec{R}_{j})}$,
where the potential between charges ($e_{j}=\pm1$) in Fourier space is
$V(\vec{q} )=(\vec{q}^{~2})^{-1}$ and $\Gamma_{V}$ is a friction coefficient. 
We use Mori's formalism \cite{beck,gotze}, to relate $\mu_{V}(\omega)$
to the correlation function $\Phi_{\rho\rho}(\vec{q},\omega)$ for the vortex
charge density $\rho(\vec{R})=\sum_{i}^{}{e_{i}\delta(\vec{R}-\vec{r}_{i})}$.
The inverse mobility of a given particle or effective viscosity function is the
sum
of the bare friction coefficient and a contribution that is related to the
forces from
all other particles :
\begin{equation}
\mu_{V}^{-1}(\omega)=\Gamma_{V}\left[ 1+
(k_{B}T)^{-1}
\sum_{\vec{q}}^{}{\left|\vec{q}V(\vec{q})\right|^{2}\Phi_{\rho\rho}
(\vec{q},\omega)} \right].
\end{equation}
\includegraphics{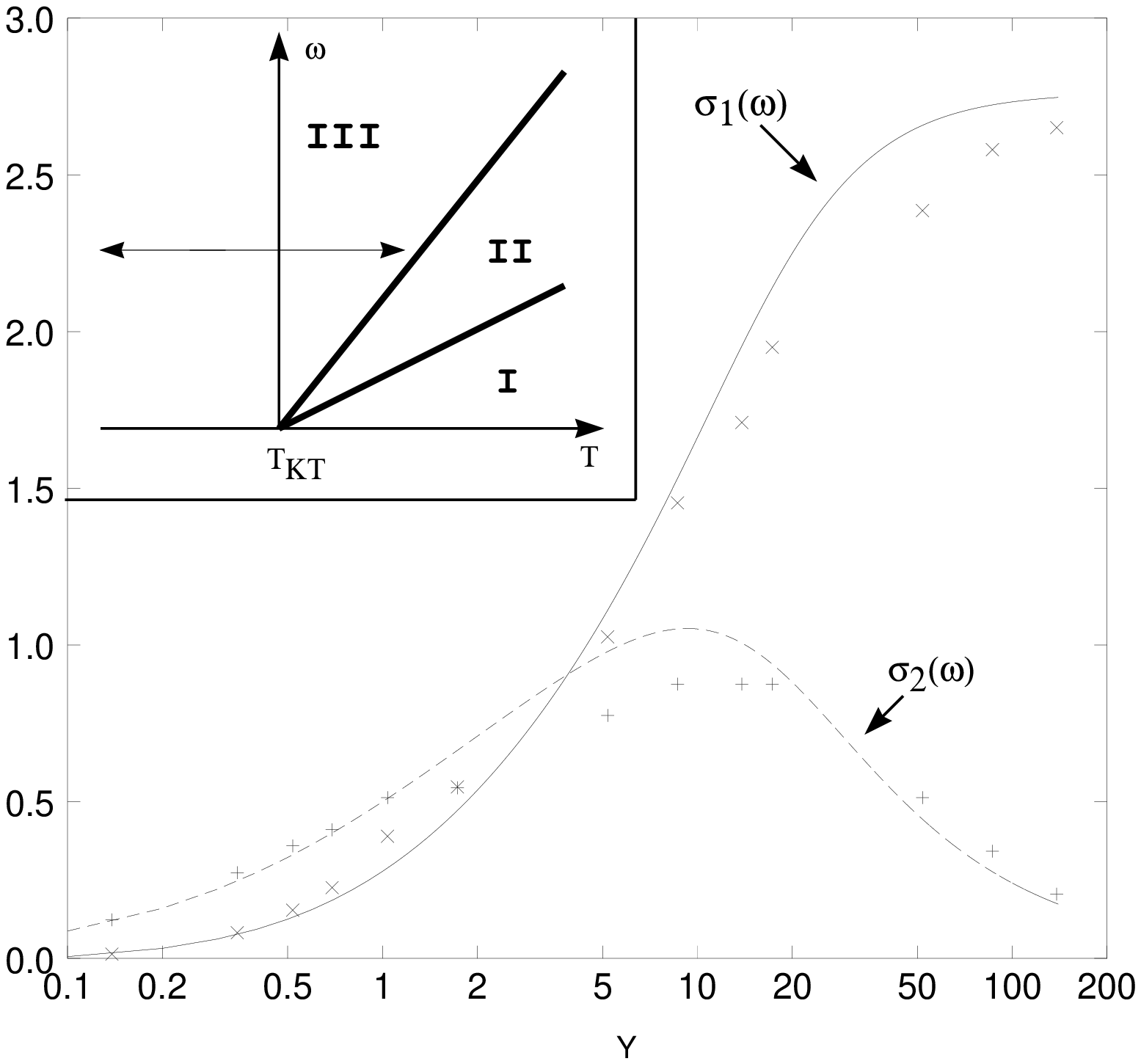}
\vglue 6.0truecm
\medskip
\begin{figure}
\protect\caption{Real and imaginary parts of the conductivity
$\sigma/\sigma_{0}\xi^{2}$
versus $Y\equiv\Gamma_{0}/\omega\xi^{2}$, in a linear-log plot, from Eqn. (5).
Here $\xi$ is the vortex screening length and 
$(\Gamma_{0}/\omega)^{1/2}$ 
the probe length $r_{\omega}$. The +, x symbols are experimental data
points \protect
[4]. Inset : regimes I, II, III in a schematic
frequency-temperature diagram, with $r_{\omega}^{-1}=\xi_{+}^{-1}(T)$ defining
the II/I boundary.}
\label{fig1}
\end{figure}
\noindent
In order to evaluate  $\Phi_{\rho\rho}(\vec{q},\omega)$ we make the usual
approximations \cite{beck} (neglect of the "Mori projector" and decoupling
higher correlations in terms of $\rho(\vec{q})$ and $n(\vec{q})$, the number
density Fourier component). One obtains :
\begin{equation}
\Phi_{\rho\rho}(\vec{q},\omega)=
{S_{\rho}(\vec{q}) \over i\omega+{k_{B}T(q^{2}+\xi^{-2}) \over
\mu_{V}(\omega)}}.
\end{equation}
Here, $S_{\rho}(\vec{q})$ is the static (charge) structure factor. Eqns. (8) and
(9) determine $\mu_{V}(\omega)$ self-consistently, but we solve to leading
order, replacing $\mu_{V}^{-1}$ in Eqn. (9) by the zeroth order $\Gamma_{V}$.
An approximate form is chosen for
$S_{\rho}(\vec{q})=
{k_{B}T \vec{q}^{2} \over 2\pi
n_{0}e^{2}}\Theta(\left|\vec{q_{1}}\right|-\left|\vec{q}\right|)+
\Theta(\left|\vec{q}\right|-\left|\vec{q_{1}}\right|)$ ($n_{0}$ being the total
density
of the charges and $\left|\vec{q_{1}}\right|$ a cut-off, corresponding
 to the first maximum in $S_{\rho}(\vec{q})$). We find :
\begin{equation}
\mu_{V}^{-1}(\omega)=
\Gamma_{V}\left[ 1+{\pi J \over k_{B}T} \ln{\left(1+{\vec{q_{1}}^{2}
\over i\omega  \Gamma_{V}+\xi^{-2}}\right)}\right].
\end{equation}
neglecting at first the second term of $S_{\rho}(\vec{q})$.
The Coulomb-gas dielectric function of the system is related to the charge
mobility
$\mu_{V}$ and to the bound-pair part of the dielectric function by \cite{beck}
$\varepsilon(\omega)=\varepsilon_{B}+i e^{2} n_{0}\mu_{V}(\omega)/\omega$.
\\
\indent
We now present the results.  Fig. 1 shows, from Eqn. (5), 
 $\sigma_{2}/\sigma_{0}\xi^{2}$,
as well  as $\sigma_{1}/\sigma_{0}\xi^{2}$ (that for small $\omega$ is essentially
$\mu_{V}^{-1}(\omega)$ the inverse vortex mobility, or vortex viscosity)
versus the logarithmic scaled frequency or temperature variable,
$\ln{Y^{-1}}$.  The experimental data \cite{theron}
have been obtained for field-induced free vortices ($\xi^{-1}=\xi^{-1}_{D}(f) \neq 0$)
for which regime III is absent and
$g(l)=(1+(a/\xi)^{2})^{-1}$, $\forall a$. Thus the
coupling $K_{l}$ should scale to zero for $l \rightarrow \infty $
 (corresponding to $T>T_{KT}$ in the zero field case). 
We use
the simple form $K_{l}=K_{0}\Theta(l_{max}-l)$, and use $l_{max}$
as fitting parameter. A good fit is obtained for $l_{max}=4.81$, that
is on the order of the l-value for which the linearized scaling equations 
\cite{thouless} yield a vanishing $K_{l}$.
$\sigma_{1}(\omega)$ clearly matches Drude behaviour for regime I, $Y\gg1$. 
The "intermediate" regime, $Y{\ \lower-1.2pt\vbox{\hbox{\rlap{$<$}\lower5pt\vbox{\hbox{$\sim$}}}}\ }
1$, with MP dependence $\sim \ln{Y}$ is seen to be fairly large.
The experimental \cite{theron} data points for SNS arrays, shown in
Figs. 1,2,  fall in regimes I
and II.
Very low frequency data are not unequivocal and are not shown.
 Typically \cite{theron}, $\Gamma_{0}\sim$ 300 Hz, $\omega$ varies from $\sim$10
Hz to $\sim$ 10 kHz, for $R_{J}\sim$ 2m$\Omega$ (SNS arrays) and $I_{J}\sim$100 nA,
and $\xi(f)\sim 3.1$ for $f=0.001$, so $Y$ goes from $\sim$ 0.1 to
$\sim$ 200. \\
\includegraphics{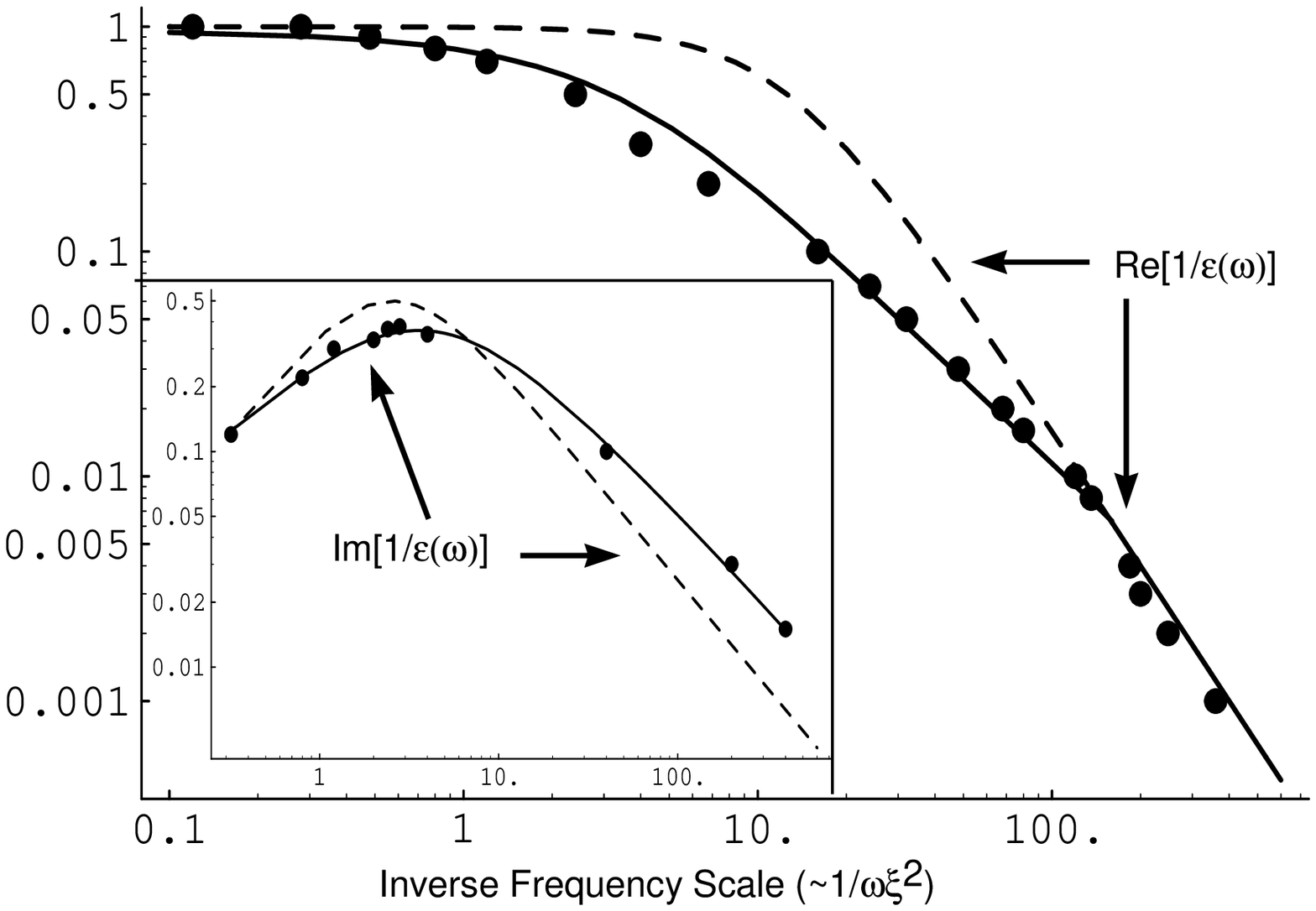}
\vglue 6.0truecm
\medskip
\begin{figure}
\caption{Real and, in the inset, imaginary parts of
$1/\varepsilon(\omega) $ versus an inverse frequency
scale $\sim Y$, in a log-log plot. The constant-mobility
($\mu_{V}^{-1}(\omega)=\Gamma_{V}$) Drude limit is
represented by the dashed line, while the solid line
is the result of our calculation b). The experimental
results of Ref. [4] are given by the dots.}
\label{fig2}
\end{figure}
\noindent
Fig. 2 shows the dielectric function $\varepsilon(\omega)$
(obtained through Eqns. (8)
and (9), by using the full form of $S_{\rho}(\vec{p})$) as a function of
Y. One again clearly recognizes two-frequency regimes (I and II), separated
by a crossover frequency $\omega_{cross}\approx(n_{0}\xi^{2})^{-1}$.
 For $\omega>\omega_{cross}$, $\Re\left(1/\varepsilon(\omega) \right)$ varies like
$\left|\omega\right|$ as in Minnhagen's regime II, whereas for
$\omega<\omega_{cross}$, it varies like 
$\omega^{2}$, as in Drude's regime I. Moreover, we
have verified for both methods, that the ratio $R_{\sigma}$ at $Y=1$
varies between the Drude ($R_{\sigma}=1$) and MP ($R_{\sigma}=2/\pi$)
 signatures \cite{capezzali}.  \\
\noindent
Regime III is not reached for the $Y$-values shown in Fig. 1, but
for SIS arrays, $R_{J}$ is orders of magnitude higher, so 
the critical behaviour might be more clearly seen. Below $T_{KT}$,
or more generally, for $Y\ll 1$, one 
has effective damping coefficients
 $\eta_{V}\sim\sigma_{1}$ due to bound pairs, rather than free-vortex
 inverse mobilities, and $\sigma_{1}\sim 1/\omega l_{\omega}^{2}$. This is
consistent with simulations
of driven vortices : there is a velocity-dependent viscosity coefficient,
decreasing for larger velocities \cite{hagenaars}. Larger oscillating pairs,
 probed at lower $\omega$, are more sluggish, since they move in a
logarithmically interacting viscous medium of smaller pairs. \\
In conclusion, we have proposed a regime interpretation of
anomalous vortex dynamics, based on the ratio of the frequency-
dependent probe scale, and the screening length. Both Drude
and anomalous vortex dynamics emerge in different regimes,
from calculations of the dynamic JJA conductivity and the
vortex mobility. This reconciles different results, supports 
postulated conductivity scaling, and
indicates further dynamical avenues to be explored, in simulations
and experiments. \\
It is a pleasure to thank D. Bormann, P. Martinoli and
P. Minnhagen for useful conversations, and P. Martinoli for reading the
manuscript.


\begin{references}
\vspace*{-1.75truecm}
\bibitem{fisher} D.S. Fisher et al., Phys. Rev. {\bf 43}, 130 (1991);
\bibitem{dorsey} A.T. Dorsey, Phys. Rev. {\bf 43}, 7575 (1991); A.T. Dorsey et al.,
Phys. Rev. {\bf 45}, 523 (1992)
\bibitem{minnhagen2} P. Minnhagen et al., Phys. Rev. Lett. {\bf 74}, 3672 (1995).
\bibitem{theron} R. Th\'{e}ron et al., Phys. Rev. Lett. {\bf 71}, 1246 (1993).
\bibitem{thouless}  J.M. Kosterlitz and D.J. Thouless, J. Phys. C
 {\bf 6}, 1181 (1973); J. Phys. C {\bf 7}, 1046 (1974).
\bibitem{mooij}  J.E. Mooij and G. Sch\"{o}n, eds., Physica B+C {\bf 157},
 (1987); H. A. Cerdeira and S.R. Shenoy, eds., Physica B {\bf 222} (1996).
\bibitem{minnhagen} P. Minnhagen, Rev. Mod. Phys. {\bf 59}, 1001 (1987);
M. Wallin, Phys. Rev. B {\bf 41}, 6575 (1990);
P. Minnhagen and O. Westman, Physica C{\bf 220}, 347 (1994);
J. Houlrik et al., Phys. Rev. B {\bf50}, 3953 (1994).
\bibitem{capezzali} H. Beck, M. Capezzali, S.R. Shenoy (unpublished).
\bibitem{ahns} V. Ambegaokar et al., Phys. Rev. B {\bf 21}, 1806 (1980).
\bibitem{shenoy} S.R. Shenoy, J. Phys. C {\bf 18}, 5143 and 5163 (1985); 
J. Phys. C {\bf 20}, 2479 (1987).
\bibitem{beck} H. Beck, Phys. Rev. B {\bf 49}, 6153 (1994).
\bibitem{korshunov} S.E. Korshunov, Phys. Rev. B {\bf 50}, 13616 (1994).
\bibitem{note3} For SNS arrays, in addition to intergrain normal currents
described by effective damping $\nu_{0}$, there can be grain-to-
substrate losses of normal current, incorporated in a generalized
TCC dynamics as a local current term $(1/\nu_{1})\dot{\theta}_{i}$
 in Eqn. (2). Here, the local damping rate $\nu_{1}$ is scale-
independent. However, following the argument \cite{shenoy}
through, there is, from the $(1/\nu_{1})\dot{\theta}_{i}$ term,
a decay rate in Fourier space $\nu_{1}q^{2}$ that for  $q^{-1}$
 in $a$, $a+da$ scales as $\nu_{1}/a^{2}$, like
$\nu_{0}\rightarrow \nu_{0}/a^{2}$. Thus
SIS ($\nu_{0}$ only) and SNS ($\nu_{0}$, $\nu_{1}$)
arrays should have the same Coulomb-gas dynamics,
as supported by calculation b).
\bibitem{teitel} K.K. Mon and S.Teitel, Phys Rev. Lett. {\bf 62}, 673 (1989).
\bibitem{gotze} W. G\"{o}tze, Phil. Mag. B {\bf 43}, 219 (1981).
\bibitem{hagenaars} T. Hagenaars et al., Phys. Rev. B {\bf 50}, 1143 (1994).

\end{references}
\end{document}